\documentclass[preprintnumbers, pra, showpacs ,floatfix,twocolumn,
preprintnumbers, letterpaper, superscriptaddress]{revtex4-1}
\usepackage{amsfonts}
\usepackage{amsmath}
\usepackage{amssymb,epsf}
\usepackage{latexsym}
\usepackage{graphicx,epsfig}
\usepackage{amssymb}
\usepackage{subfigure}
\usepackage[colorlinks=true,citecolor=blue,linkcolor=blue,urlcolor=black]{hyperref}
\usepackage{epstopdf}
\usepackage{color,soul}

\def\be{\begin{equation}}
\def\ee{\end{equation}}
\def\ba{\begin{eqnarray}}
\def\ea{\end{eqnarray}}

\begin{document}
\title{Global entanglement in a topological quantum phase transition }
\author{Elahe Samimi}
\email{elh.samimi@gmail.com}
\affiliation{Department of Physics, School of Science, Shiraz University, Shiraz 71946-84795, Iran}

\author{Mohammad Hossein Zarei}
\email{mzarei92@shirazu.ac.ir}
\affiliation{Department of Physics, School of Science, Shiraz University, Shiraz 71946-84795, Iran}

\author{Afshin Montakhab}
\email{motomonty@gmail.com}
\affiliation{Department of Physics, School of Science, Shiraz University, Shiraz 71946-84795, Iran}

\begin{abstract}
A useful approach to characterize and identify quantum phase
transitions lies in the concept of multipartite entanglement. In
this paper, we consider well-known measures of multipartite
(global) entanglement, i.e., average linear entropy of one-qubit
and two-qubit reduced density matrices, in order to study
topological quantum phase transition (TQPT) in the Kitaev Toric
code Hamiltonian with a nonlinear perturbation. We provide an
$exact$ mapping from aforementioned measures in the above model to internal energy and energy-energy correlations in the classical Ising model. Accordingly, we find that the global entanglement shows a continuous and sharp transition from a maximum value in the topological phase to zero in the magnetized phase in a sense that its first-order derivative diverges at the transition point. In this regard, we conclude that not only can the global entanglement 
serve as a reasonable tool to probe quantum criticality at TQPTs, but it also can reveal the highly entangled nature of topological phases. Furthermore, we also introduce a conditional version of global entanglement which becomes maximum at the critical point. Therefore, regarding a general expectation that multipartite entanglement reaches maximum value at the critical point of
quantum many-body systems, our result proposes that the conditional global entanglement can be a good measure of multipartite entanglement in TQPTs.
\end{abstract}
 \maketitle
\section{Introduction}
Entanglement \cite{entangled,Horodecki} as the hallmark of quantum physics plays a pivotal role in describing various quantum phenomena. In particular, it provides a general framework to study quantum phase transitions (QPTs) both theoretically and experimentally \cite{wires,susc,view,zhao,vidal,roman}.
Nevertheless, the behavior of entanglement at critical
quantum 
many-body systems is relatively puzzling. While the divergence
of correlation length implies that total entanglement might
be maximum at the QPT point, it does not happen for pairwise entanglement
\cite{simple,fazio}. It has been argued that monogamy property
puts a limit on the amount of distributed pairwise entanglement
at quantum criticality \cite{simple}. This seemingly reinforces
the notion that it is the multipartite entanglement which should
be maximum at quantum criticality \cite{bayat}.

On the other hand, regarding the complexity of entanglement in many-body systems, there is not a single measure for multipartite entanglement \cite{Horodecki}. This in turn highlights the very real need for considering which multipartite entanglement measures can characterize and quantify entanglement unambiguously \cite{comput}. One of the most well-known measures of multipartite entanglement is global entanglement ($GE$) \cite{meyer}, which is in fact the average linear entropy of one-qubit reduced density matrices \cite{genuine,Lak}. Since $GE$ captures all quantum correlations in the system, it is a suitable tool for characterizing QPTs in quantum many-body systems \cite{asadian,montakhab,rad18,localization,vimal18}. However, the behavior of $GE$ in a critical system is also affected by symmetries and finite-size effects. For example, while $GE$ becomes maximum at the critical point in the thermodynamic limit \cite{genuine,operational,rad18}, for a finite system it is a monotonic function of coupling and its first-order derivative diverges logarithmically with system size \cite{montakhab,rad18}. Accordingly, it seems that the symmetry-breaking mechanism \cite{Landau}, which happens in the thermodynamic limit, plays the key role in maximization of $GE$ at the critical point \cite{effect}. However, in \cite{lipkin}, the author has studied a specific model and  showed that $GE$ does not reach a maximum value at the critical point in spite of the existence of a symmetry-breaking mechanism.

Furthermore, there are other kinds of QPTs which cannot be described through a symmetry-breaking mechanism, i.e., TQPTs \cite{topological,zoo}. There is no local order parameter for characterizing topological phases and instead they are highly entangled states with long-range entanglement which can be characterized by topological entanglement entropy \cite{detecting,topological2,nat,yao2010}. Such a long-range quantum correlation particularly leads to a robust degeneracy which has important applications for quantum information processing tasks \cite{memory}. In spite of having a different kind of correlation, the theory of critical phenomena is still applicable to TQPTs such that it is possible to define critical exponents, scaling relations, and finite-size effects \cite{Multicritical, Multi-critical,Casimir,finite}. In this regard, it is an important task to study how long-range entanglement plays a role in the critical behavior of the system and specifically how it affects the behavior of total quantum correlation measured by $GE$. Furthermore, due to the lack of a comprehensive mechanism for TQPTs, considering the behavior of multipartite entanglement can lead to a better understanding of the mechanism of such phase transitions \cite{diagonal,smerzi}.

Here, we study TQPT in a perturbed version of the Kitaev Toric code model \cite{Kitaev2003}. The Toric code has been studied in the presence of different types of perturbations where TQPT points are also important as a measure of the robustness of the topological phase against perturbations \cite{string,robustness2,robustness,ham,thom,castel}. The critical behavior in different quantities including ground-state fidelity \cite{zanardi,hamma}, quantum discord \cite{chen}, and quantum Fisher information \cite{Zhang} has been studied. While there are different approaches such as tensor network methods for studying TQPTs \cite{schotte}, mapping to statistical mechanical models has also been known as a simple and rich method \cite{jahromi,afshin}. In particular, here we consider a Toric code model in the presence of a nonlinear perturbation, where a simple correspondence to a classical Ising model is established \cite{castel}. We consider $GE$ and generalized global entanglement ($\widetilde{GE}$) to find what quantities in a classical Ising model they are mapped to and then to illustrate how they can detect criticality. First, by $exact$ analytical calculations, we reach simple mathematical formulas that relate $GE$ and $\widetilde{GE}$ to internal energy and energy-energy correlations of a 2D classical Ising model, respectively. Then, using such analytical relations together with numerical simulations of a classical Ising model, we show that both $GE$ and $\widetilde{GE}$ are decreasing monotonic functions of coupling and criticality can be marked by the divergence (maximum) of the first-order derivative of $GE$ and $\widetilde{GE}$ at the thermodynamic limit (for a finite-size quantum system). This result supports a possible connection between the maximization of global entanglement and the symmetry-breaking mechanism. Finally, we look for a good measure of multipartite entanglement for the model in a sense that it peaks at the critical point. To this end, we use the concept of quantum conditional entropy \cite{Nielsen} and show that there is a suitable measure in the form of $\widetilde{GE}-GE$, which peaks at the critical point. It also reveals the role of long-range entanglement in the monotonic behavior of global entanglement in the model under consideration.

The paper is organized as follows. In Sec. \ref{sec1}, we define the model which is the Kitaev Hamiltonian in the presence of a nonlinear perturbation and give an overview of some properties of topological order. We also explain how the quantum model maps to the classical Ising model. In Sec. \ref{sec2}, we define $GE$ and $\widetilde{GE}$ and present our numerical results about how these quantities behave in TQPT. Finally, in Sec. \ref{sec3}, we introduce a new parameter, namely conditional global entanglement, which is equal to the difference between $\widetilde{GE}$ and $GE$ and numerically show that it is maximum at the critical point. We also provide analytical explanation to prove our point.

\section{The model}\label{sec1}
\begin{figure}[t]
\centering
\includegraphics[width=7.5cm,height=5cm,angle=0]{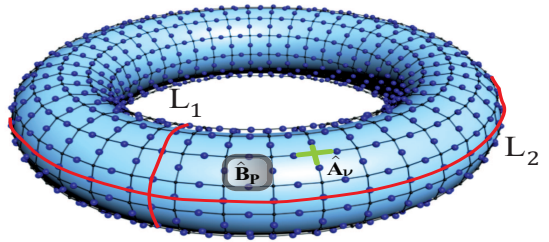}
\caption{(Color online) 2D square lattice with periodic boundary condition. Qubits exist on the edges. Two nontrivial loops $L_{1}$ and $L_{2}$ together with $\hat A_{v}$ (star operator) and $\hat B_{p}$ (plaquette operator) are shown.} \label{fig:1a}
\end{figure}

We consider the Kitaev Toric code Hamiltonian with a nonlinear perturbation which shows TQPT \cite{castel}. The model is defined on a square lattice with a periodic boundary condition in which spin-$1/2$ particles live on the edges (see Fig. \ref{fig:1a}). The Hamiltonian is given by

\begin{equation}\label{hamilton}
\hat H=-\sum_v \hat A_v -\sum_p \hat B_p +\sum_v e^{-\beta\sum_{i\in v }\hat{\sigma}_i^z},
\end{equation}
where $ \beta >0$
is a coupling constant. $\hat A_v$ and $\hat B_p$ are the star and plaquette operators, respectively, defined by
\\
\begin{equation}\label{sta}
\hat A_v =\prod_{i\in v} \hat{\sigma}_i^x~~,~~\hat B_p =\prod_{i\in \partial p} \hat{\sigma}_i^z.
\end{equation}
\\
$\hat{\sigma}_i^z$ and $\hat{\sigma}_i^x$ are the Pauli operators. As shown in Fig. \ref{fig:1a} $\hat A_v$  acts on the four qubits connected to the vertex $v$ and $\hat B_p$ acts on the four qubits around the plaquette p.

\begin{figure}[t]
\centering
\includegraphics[width=8cm,height=7cm,angle=0]{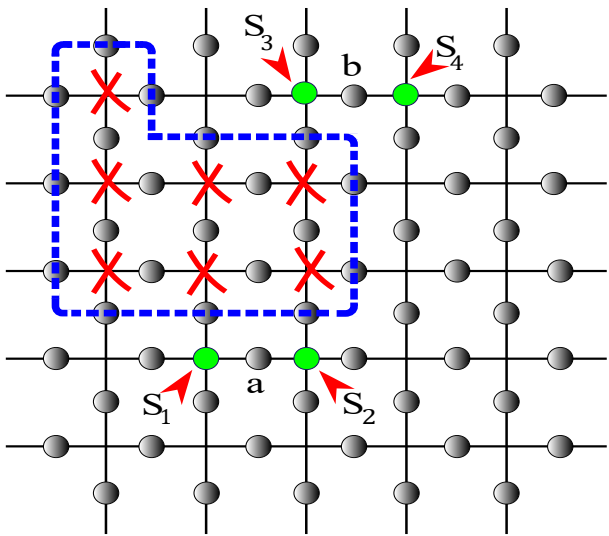}
\caption{(Color online) 2D square lattice with periodic boundary condition. Despite qubits (in the quantum model) which exist on the edges (like qubits $a$ and $b$), classical spins (in the classical Ising model) such as $S_{1}$, $S_{2}$, $S_{3}$, and $S_{4}$ are on the vertices of the square lattice. The closed loop with dashed blue line corresponds to a particular $g$. The product of some $\hat A_{v}$'s, marked by multiplication signs, has created the loop.} \label{fig:1b}
\end{figure}

Note that when $\beta=0$ the Hamiltonian reduces to the Kitaev model with a trivial constant. Since all star and plaquette operators commute with each other, it is quite straightforward to prove that one of the ground state wave functions of the Kitaev Hamiltonian, up to a normalization factor, is in the following form:
\begin{equation}\label{gs1}
|GS\rangle=\prod_{v}(1+\hat A_{v})|0\rangle ^{\otimes{N}},
\end{equation}
where $N$ is the total number of qubits and $|0\rangle ^{\otimes{N}}$ denotes the fully magnetized state wherein the eigenvalues of all $\hat{\sigma}_i^z$ become $+1$. Equation (\ref{gs1}) suggests that the ground state is a superposition state obtained by summing over all possible products of star operators which act on $|0\rangle ^{\otimes{N}}$. In terms of stabilizer formalism, $\hat A_v$'s are generators of an Abelian group $G$ with $2^{\frac{N}{2}-1}$ elements in the following form:
\begin{equation}\label{stab}
g_{\{r_{1}, r_{2}, ..., r_{N/2}\}}=\hat A_1^{r_1}. \hat A_2^{r_2}. ...  . \hat A_{\frac{N}{2}-1}^{r_{\frac{N}{2}-1}},
\end{equation}
where $r_i=\{0,1\}$ and $N/2$ is the total number of vertices. Such a product of star operators can be represented by a loop configuration, simply because each $\hat A_v$ forms a loop in the dual lattice. Figure \ref{fig:1b} shows a particular loop operator achieved by applying the product of some $\hat A_v$'s, denoted by multiplication signs. Therefore, the wave function in Eq. (\ref{gs1}) is a loop condensed state which has a special kind of order called topological order. It is also known that, due to the nontrivial topology of the lattice, there is a fourfold degenerate subspace constructed by applying two noncontractible loop operators, see. Fig. \ref{fig:1a}, to the ground state in Eq. (\ref{gs1}). The dependence of degeneracy to topology as well as the nonlocal nature of the ground states are common properties of topological phases. 
 
Now, let us turn back to the perturbed Hamiltonian and consider $\beta \rightarrow \infty$. In this case, the ground state becomes the fully magnetized state $|0\rangle^{\otimes N}$, which is topologically trivial. Therefore, one would expect that by tuning the parameter $\beta$ from zero to $\infty$,  the ground state of Eq. (\ref{hamilton}) shows a TQPT from a topological phase to a magnetized phase at a quantum critical point $\beta^*$.

Fortunately, the exact ground state of Eq. (\ref{hamilton}) has been obtained analytically \cite{castel} and it can be written as
\begin{equation}\label{GS}
|GS(\beta)\rangle=\frac{1}{\sqrt{Z(\beta)}}\sum_{g\in{G}} e^{\frac{\beta}{2}\sum_i \sigma_i^z(g)} g|0\rangle^{\otimes N},
\end{equation}
where $g \in G$ refers to loop operators defined in Eq. (\ref{stab}). $ Z(\beta)=\sum_{g\in{G}}e^{\beta\sum_i \sigma_i^z(g)}$ and $\sigma_i^z(g)=-1$ $(\sigma_i^z(g)=+1)$ if the qubit $i$ and the loop $g$ have an (do not have any) intersection.

Next, let us pay attention to a special feature of Eq. (\ref{GS}) which is extremely relevant to our study. According to \cite{castel}, there is an exact correspondence between the partition function of a 2D classical Ising model, ruled by the Hamiltonian $H=-J\sum_{\langle k {k}'\rangle} S_{k}S_{{k}'}$
 where $S_{k}$ and $S_{{k}'}$ refer to classical spins $k$ and ${k}'$ in the Ising model, and the normalization factor in Eq. (\ref{GS}). This correspondence originates from the fact that probability amplitudes in Eq. (\ref{GS}) are simply related to the Boltzmann weights of spin configurations in the classical Ising model. The following lines of this section seek to explain more about the general concept of this established quantum-classical relation.

Consider the Ising model with classical spins (like $S_{1}$, $S_{2}$, $S_{3}$, and $S_{4}$ in Fig. \ref{fig:1b}) attached on the vertices of the square lattice with periodic boundary condition. In the low-temperature expansion of the 2D classical Ising model, each spin configuration corresponds to a closed loop pattern in the dual lattice which separates upward spins from downward spins by joining lines that cross edges with unlike ends \cite{pathria}.  It is easy to graphically reach the conclusion that spin configurations in the Ising model produce exactly the same closed loop patterns in the dual lattice as generators in the quantum model. In this regard, $\sum_g$ in the $Z(\beta)$ is replaced by $\sum_{C}$, which is the sum over Ising spin configurations denoted by $C$. However, since there are two spin configurations corresponding to each loop configuration,  the exact transformation becomes $\sum_g = 1/2\sum_{C}$. Furthermore, the value of $\sigma_i^z(g)$ can be determined solely by two Ising spins attached to the end points of the $i^{th}$ edge. For example, quantum configurations in which $g$ crosses qubit $i$ and $\sigma_i^z(g)=-1$, correspond to classical configurations where the two ends of the $i^{th}$ edge have the opposite directions.  Therefore, we can write $\sigma_i^z(g) = S_{k}S_{k'}(C)$ and the summation over qubits $\sum_i$ in the quantum model can be replaced by the summation over nearest-neighbor spins in the classical model, i.e., $\sum_i = \sum_{\langle k k'\rangle}$. Regarding these relations, the normalization factor is written in the form of
\begin{equation}\label{Z}
Z(\beta)=\frac{1}{2}\sum_{C}e^{\beta\sum_{\langle k k'\rangle}S_{k}S_{k'}}.
\end{equation}
By considering the summation in the right-hand side of Eq. (\ref{Z}) as the partition function of the Ising model, we can immediately deduce that the parameter $\beta$ in the quantum model corresponds to $J/k_{B}T$ in the Ising model, where $J$, $k_{B}$, and $T$ are coupling constant of interactions, Boltzmann constant, and temperature, respectively. As a matter of simplicity, we have assigned the value of $1$ to $J$ and $k_{B}$.

Making use of the quantum-classical mapping, it is shown that the fidelity of the quantum model directly relates to the heat capacity of the classical Ising model. As a result, a singularity in the heat capacity is in conjunction with a corresponding singularity in the quantum model which is a clear sign of QPT \cite{zanardi}. Furthermore, as evidenced by \cite{zarei20,zarei18,zarei21,van}, there exists some other examples which show this kind of quantum-classical correspondence.

\section{global entanglement}\label{sec2}
In the previous section, it became clear that the quantum model defined in Eq. (\ref{hamilton}) shows a TQPT at $\beta^*$ corresponding to the classical phase-transition temperature of the 2D Ising model. It seems that using the above quantum-classical mapping, one might be able to consider different important quantities for the perturbed Toric code model by finding the corresponding quantity in the 2D Ising model. Here we consider $GE$ proposed by Meyer and Wallach \cite{meyer} as a measure of multipartite entanglement and ask if it can diagnose criticality in the perturbed Toric code model. 

In order to study $GE$, notice that, for an $N$-qubit system, it is defined as the mean linear entropy of one-qubit reduced density matrices of the system in the following form \cite{Lak,genuine}:
\begin{equation}\label{GE}
GE =2(1-\frac{1}{N}\sum_{i=1}^N Tr(\hat \rho_{i}^2)),
\end{equation}
where $\hat \rho_{i}$ is the reduced density matrix corresponding to qubit $i$. In addition, one can also use a generalization of $GE$ as the average linear entropy of two-qubit reduced density matrices in the form of \cite{operational}
\begin{equation}\label{GGE}
\widetilde{GE} =\frac{4}{3}(1-\frac{2}{N(N-1)}\sum_{(ij)} Tr(\hat \rho_{ij}^2)),
\end{equation}
where $\hat \rho_{ij}$ refers to the reduced density matrix corresponding to qubits $(ij)$. $ Tr(\hat\rho_i ^2)$ in Eq. (\ref{GE}) measures the degree of purity of the state $\hat\rho_i$ and $2(1- Tr(\hat\rho_i ^2))$ is the linear entropy of $\hat\rho_i$ which characterizes entanglement between the qubit $i$ and other qubits of the system. With the same explanation, $\widetilde{GE}$ can characterize average entanglement between two-qubit reduced density matrices and the rest of the system. Factors $2$ and $4/3$ in Eq. (\ref{GE}) and Eq. (\ref{GGE}) normalize the maximum value of entanglement to $1$.

In order to compute $GE$ and $\widetilde{GE}$ for the ground state of the quantum model, we consider $|GS(\beta)\rangle \langle GS(\beta)|$ and then we find reduced density matrices $\hat\rho_{a}$ and $\hat\rho_{ab}$, where $a$ and $b$ refer to particular qubits of the system. To this end, we should trace out the rest of the system in the following forms:
\begin{equation}\label{one}
\begin{aligned}
\hat{\rho}_{a}=\frac{1}{Z}\sum_{\lbrace \alpha_{m}=0,1|m\neq{a}\rbrace}\sum_{g,g'}e^{\frac{\beta}{2}\sum_i [\sigma_i^z (g)+\sigma_i^z(g')]}
\\
\times\langle \alpha_1, \alpha_2, ..., \alpha_N|  g|0\rangle^{\otimes N} ~^{N\otimes}\langle 0|g'|\alpha_1, \alpha_2, ..., \alpha_N\rangle,
\end{aligned}
\end{equation}
\begin{equation}\label{two}
\begin{aligned}
\hat{\rho}_{ab} =\frac{1}{Z}\sum_{\lbrace \alpha_m=0,1|m\neq{a,b\rbrace}}\sum_{g,g'}e^{\frac{\beta}{2}\sum_i [\sigma_i^z (g)+\sigma_i^z(g')]}
\\
\times\langle \alpha_1, \alpha_2, ..., \alpha_N|  g|0\rangle^{\otimes N} ~^{N\otimes}\langle 0|g'|\alpha_1, \alpha_2, ..., \alpha_N\rangle.
\end{aligned}
\end{equation}
Only diagonal terms of $\hat\rho_{a}$ and $\hat\rho_{ab}$ appear, i.e., $\langle\alpha_a|\hat{\rho}_{a}|{\alpha}'_a\rangle\neq{0}$ only if $\alpha_a ={\alpha}'_a$, while $\langle\alpha_a\alpha_b|\hat{\rho}_{ab}|{\alpha}'_a{\alpha}'_b\rangle\neq{0}$ on the condition that
$\alpha_a ={\alpha}'_a$ and $\alpha_b={\alpha}'_b$. Let us explain what makes it impossible for $\hat\rho_a$ to have a non-diagonal term  $\langle\alpha_a|\hat{\rho}_{a}|{\alpha}'_a\rangle$. According to Eq. (\ref{one}) this term is proportional to
\begin{equation}\label{rone}
\begin{aligned}
&(\langle \alpha_1,\alpha_2, ..., \alpha_a, ...,\alpha_N|) (g_{1}|0\rangle^{\otimes{N}})
\\
&\times({}^{ N\otimes}\langle 0| g_{2})(|\alpha_1,\alpha_2,..., {\alpha}'_a, ...,\alpha_N\rangle),
\end{aligned}
\end{equation}
where $g_{1}|0\rangle^{\otimes{N}}$ and $g_{2}|0\rangle^{\otimes{N}}$ denote two arbitrary distinct closed loop patterns which correspond to two elements of group $G$. Therefore, only states  $|\alpha_1,\alpha_2,..., {\alpha}_a, ...,\alpha_N\rangle$ and $|\alpha_1,\alpha_2,..., {\alpha}'_a, ...,\alpha_N\rangle$ which configure closed loops can have nonzero inner products. Equation \ref{rone} implies that $\langle\alpha_a|\hat{\rho}_{a}|{\alpha}'_a\rangle\neq{0}$ if there exists two closed-loop configurations $g_{1}|0\rangle^{\otimes{N}}$ and $g_{2}|0\rangle^{\otimes{N}}$ such that they differ in just one qubit, i.e., one link. However, it is impossible to have such two different loop configurations, since with the combination of at least two loops, a new loop pattern is constructed, and not with a closed loop and an open string.

The same assertion with a similar argument holds for $\hat\rho_{ab}$. It means nondiagonal terms, wherein $\alpha_{a}\neq{{\alpha}'_a}$ or $\alpha_{b}\neq{{\alpha}'_b}$, do not exist. Regarding these facts, we obtain
\begin{equation}\label{three}
\hat\rho_{a} =\frac{1}{Z}(Z_{0}|0\rangle\langle0|+Z_{1}|1\rangle\langle1|),
\end{equation}
\begin{equation}\label{four}
\begin{aligned}
\hat\rho_{ab}&=\frac{1}{Z}(Z_{00}|00\rangle\langle00|+Z_{01}|01\rangle\langle01|)
\\
 &+\frac{1}{Z}(Z_{10}|10\rangle\langle10|+Z_{11}|11\rangle\langle11|)),
\end{aligned}
\end{equation}
where $Z_{0}/Z$ ($Z_{1}/Z$) is the sum of the squares of probability amplitudes of configurations in which the state of qubit $a$ is $|0\rangle$ ($|1\rangle$). Similarly, $Z_{00}/Z$, $Z_{01}/Z$, $Z_{10}/Z$, and $Z_{11}/Z$ are the sum of the squares of probability amplitudes of configurations where the state of qubits $|ab\rangle$ are $|00\rangle$, $|01\rangle$, $|10\rangle$, and $|11\rangle$, respectively.

Considering the mapping to the classical Ising model and the fact that probability amplitudes in Eq. (\ref{GS}) are related to Boltzmann weights of Ising spin configurations, it is possible to describe $Z_{0}$, $Z_{1}$, $Z_{00}$, $Z_{01}$, $Z_{10}$ and $Z_{11}$ according to quantities in the classical model. Hence, we can interpret $Z_{0}$ ($Z_{1}$) as the sum of Boltzmann weights of configurations in the classical Ising model, where the product of two nearest-neighbor spins like $S_{1}$  and $S_{2}$ in Fig. \ref{fig:1b} which are attached to the ends of the $a^{th}$ edge becomes $S_{1}S_{2}=1$ ($S_{1}S_{2}=-1$). With the same view, $Z_{00}$, $Z_{01}$, $Z_{10}$, and $Z_{11}$ are the sum of Boltzmann weights of configurations where spins of two edges like $(S_{1}$, $S_{2})$ and $(S_{3}$, $S_{4})$ in Fig. \ref{fig:1b} attached to the ends of $a^{th}$ and $b^{th}$ links satisfy relations $S_{1}S_{2}=1$ and $S_{3}S_{4}=1$, $S_{1}S_{2}=1$ and $S_{3}S_{4}=-1$, $S_{1}S_{2}=-1$ and $S_{3}S_{4}=1$, and
$S_{1}S_{2}=-1$ and $S_{3}S_{4}=-1$,
respectively. 
\begin{figure*}[t!]
\centering     
\subfigure[]{\label{fig:3a}\includegraphics[width=65mm, height=50mm]{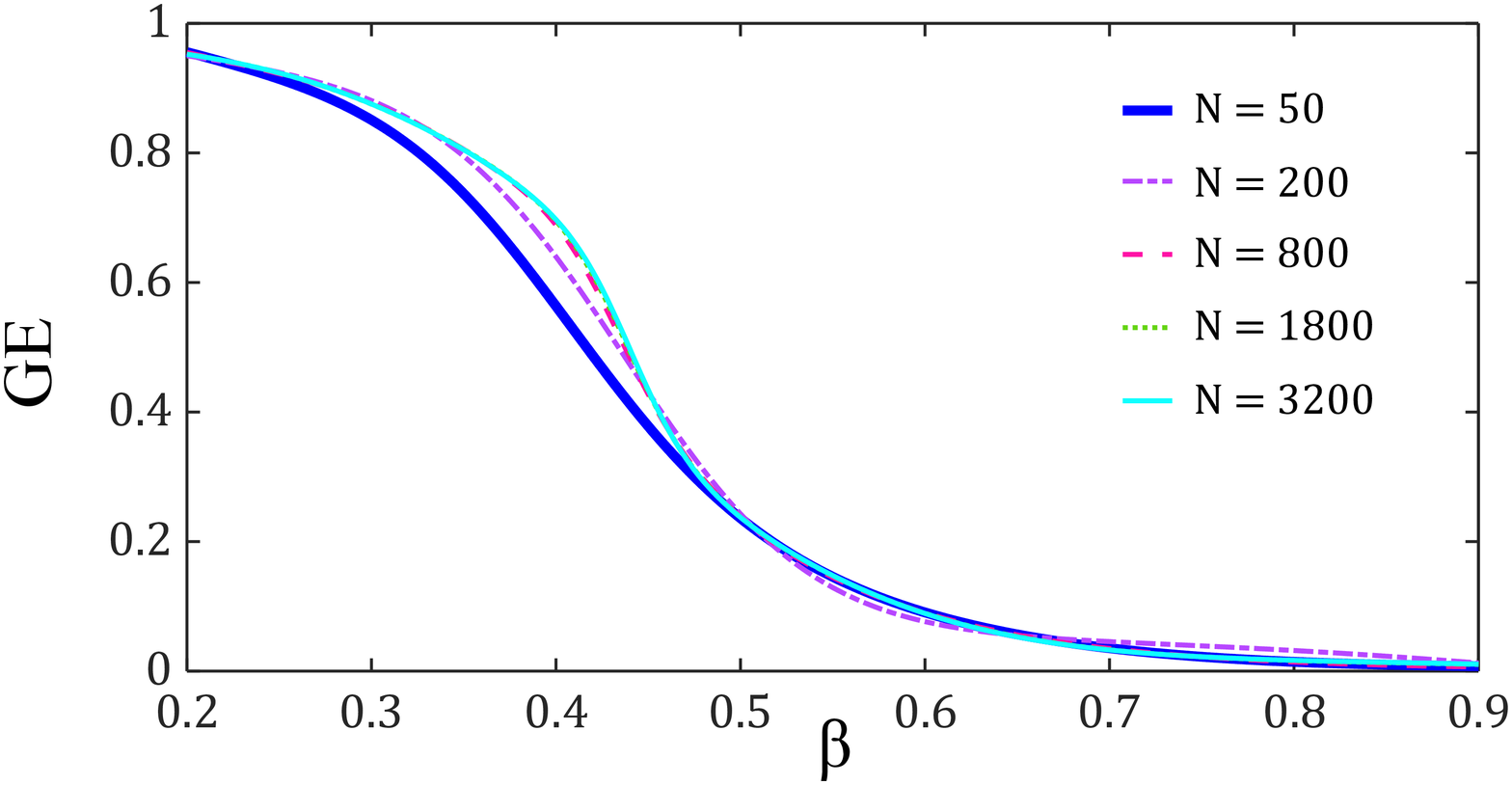}}
\subfigure[]{\label{fig:3b}\includegraphics[width=65mm,height=50mm]{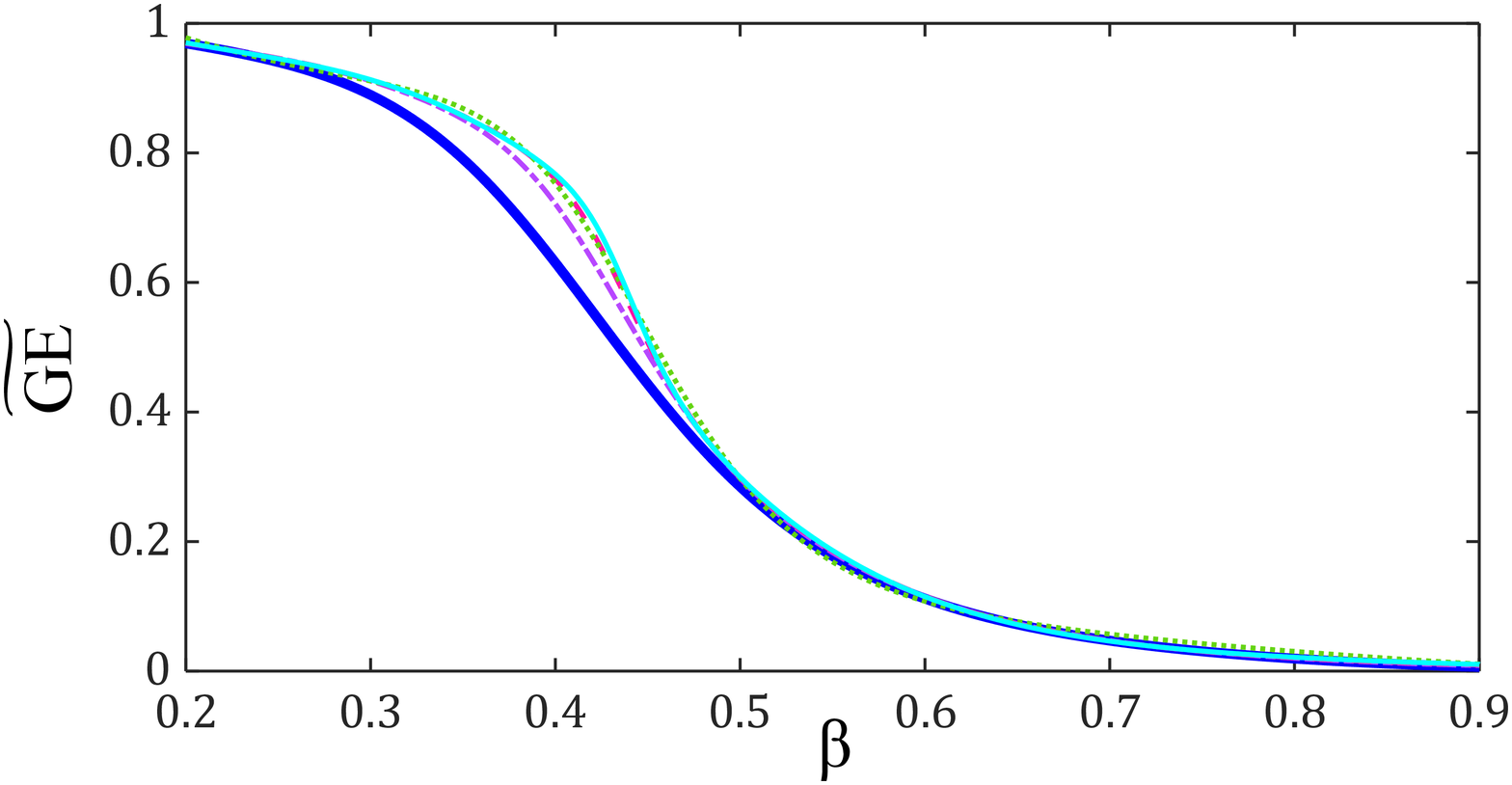}}
\subfigure[]{\label{fig:3c}\includegraphics[width=65mm,height=50mm]{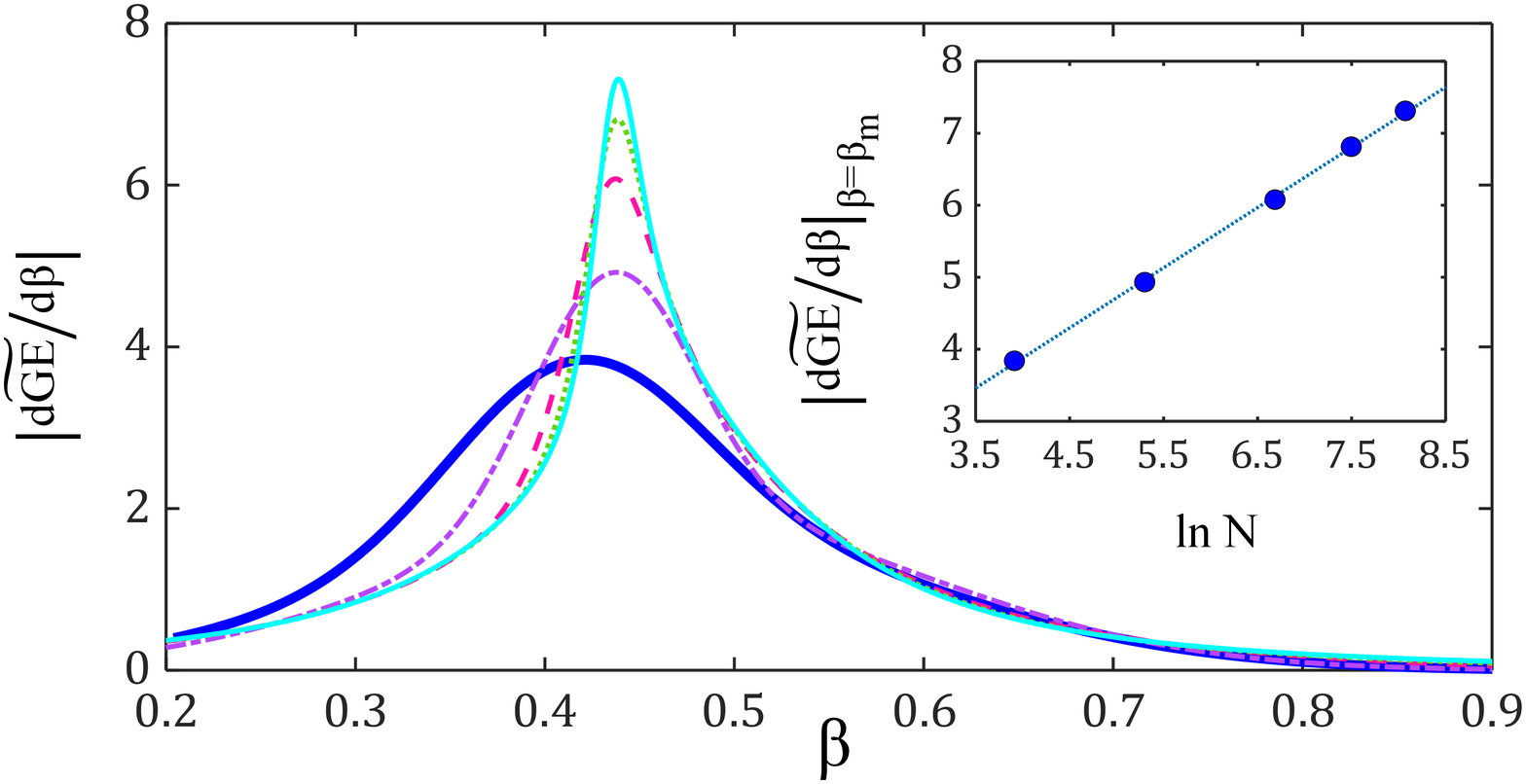}}
\subfigure[]{\label{fig:3d}\includegraphics[width=65mm,height=50mm]{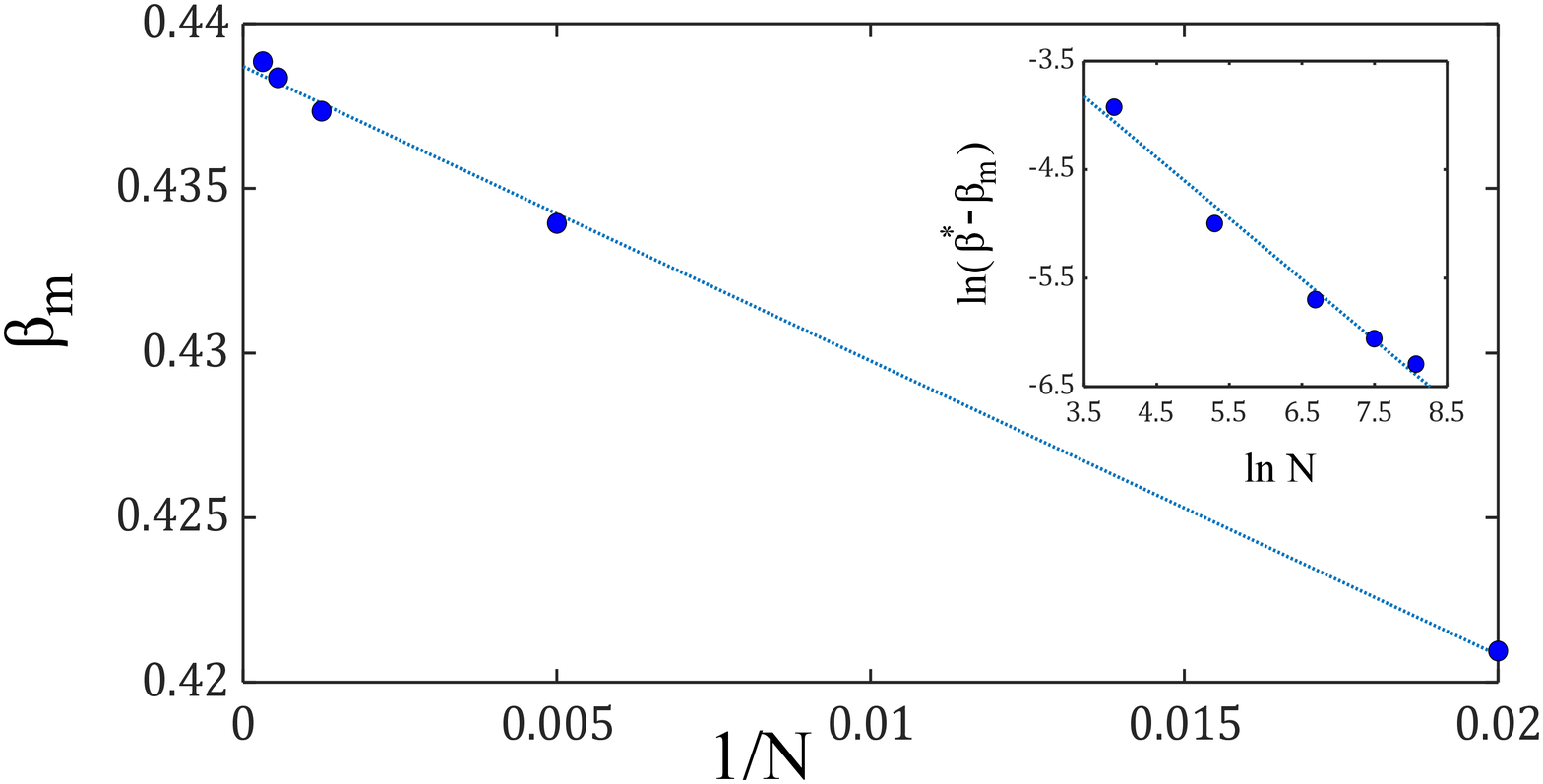}}
\caption{(Color online) (a) $GE$ (global entanglement). (b) $\widetilde{GE}$ (generalized global entanglement) vs $\beta$ (coupling) for different system sizes. (c) The first-order derivative of $\widetilde{GE}$ with respect to  $\beta$. The inset shows that the maximum diverges logarithmically $d\widetilde{GE}/d\beta|_{\beta=\beta_{m}}\approx (0.836)\ln{N}$. (d) Convergence of $\beta_{m}$ to $\beta^*$. The estimation of y interception $\beta_m({N\rightarrow\infty})\approx 0.439$. The inset suggests that $\beta_{m}$ converges to $\beta^*$ with relation $|\beta^*-\beta_{m}|\sim N^{-0.56}$. All plots are obtained by numerical simulations of the 2D classical Ising model. According to the quantum model, $N$ is the total number of qubits, while in the classical model and hence in the simulations $N/2$ is the total number of classical spins.}
\end{figure*}
Interestingly, $Z_{0}/Z$ ($Z_1 /Z$) is nothing more than the probability function in the classical Ising model when two nearest-neighbor spins have the same (opposite) directions, and hereafter denoted by $P_{s}$ ($P_o$). The same probability interpretation goes for other quantities $Z_{00}/Z$, $Z_{01}/Z$, $Z_{10}/Z$ and $Z_{11}/Z$ and henceforth identified as $P_{ss}$, $P_{so}$, $P_{os}$ and $P_{oo}$, respectively. For example, consider again two arbitrary edges $a$ and $b$; then $P_{ss}$ is the probability function when $S_{1}S_{2}=1$ and ${S_{3}S_{4}=1}$, while $P_{so}$ is the probability function when $S_{1}S_{2}=1$ and $S_{3}S_{4}=-1$. After simple calculation, Eq. (\ref{GE}) and Eq. (\ref{GGE}) can be written as
\begin{equation}\label{P1}
GE=2(1-P_{s}^{2}-P_{o}^{2}),
\end{equation}

\begin{equation}\label{P2}
\widetilde{GE}=\frac{4}{3}(1-\frac{2}{N(N-1)}\sum_{(ij)}(P_{ss}^{2}+P_{so}^{2}+P_{os}^{2}+P_{oo}^{2})).
\end{equation}
By considering the energy of link $a$, $E_{a}=-S_{1}S_{2}$, we can write
\begin{equation}\label{m}
\begin{aligned}
&\langle E_{a}\rangle=P_{o}-P_{s},
\\
&P_{s}+P_{o}=1,
\end{aligned}
\end{equation}
where $\langle E_{a}\rangle$ is the expected energy of link $a$ and the last equation is the sum of the probabilities. By exploiting the relation $\langle E_{a}\rangle^2=\langle E\rangle^2 /  N^2$, where $\langle E\rangle$ is the expectation value of total energy, together with Eq. (\ref{m}), $GE$ can be written as
\begin{equation}\label{C1}
GE=1-\frac{\langle E\rangle ^{2}}{N^2}.
\end{equation}
In the same way, we can simplify $\widetilde{GE}$. To this end, notice that for two arbitrary edges $a$ and $b$ in the Ising model we can write the following relations:
\begin{equation}\label{mm}
\begin{aligned}
&\langle E_{a}\rangle= P_{oo} + P_{os}-P_{so}- P_{ss},
\\
&\langle E_{b}\rangle= P_{oo} + P_{so}-P_{os}- P_{ss},
\\
&\langle E_{a}E_{b}\rangle= P_{ss} - P_{os}-P_{so}+P_{oo},
\\
&P_{ss}+P_{so}+P_{os}+ P_{oo}=1.
\end{aligned}
\end{equation}
Therefore, by using Eq. (\ref{P2}), $\widetilde{GE}$ can be written in the following form:
\begin{equation}\label{C2}
\widetilde{GE}=1-\frac{2}{3}\frac{\langle E\rangle ^{2}}{N^2}-\frac{2}{3N(N-1)}\sum_{(ij)}\langle E_{i}E_{j}\rangle^{2}.
\end{equation}

Equation (\ref{C1}) and Eq. (\ref{C2}) describe exact mappings between $GE$, $\widetilde{GE}$ in the quantum model and internal energy, as well as energy-energy correlations in the classical Ising model. In particular, if we calculate the derivative of Eq. \ref{C1} with respect to $\beta$, we obtain
\begin{equation}\label{dGE}
\frac{dGE}{d\beta}=-2\frac{\langle E\rangle}{N^2}\frac{d\langle E\rangle}{d\beta}.
\end{equation}
$d\langle E\rangle/d\beta$ is proportional to the heat capacity of a classical Ising model and therefore diverges at the critical point. Accordingly, the first-order derivative of $GE$ diverges at $\beta^{*}$ and consequently it can be regarded as a useful indicator of quantum criticality for the model under consideration. Considering such a mapping to classical thermodynamic quantities, we are also able to numerically calculate $GE$ and $\widetilde{GE}$ in the quantum model by direct simulation of the classical Ising model.

We have plotted $GE$ and $\widetilde{GE}$ vs $\beta$ for several system sizes. As shown in Fig. \ref{fig:3a} and Fig. \ref{fig:3b}, both $GE$ and $\widetilde{GE}$ decrease as a function of $\beta$ and in Fig. \ref{fig:3c}, $d\widetilde{GE}/d\beta$ peaks at $\beta_{m}$. The inset in Fig. \ref{fig:3c} shows that the peak of $|d\widetilde{GE}/d\beta|$ diverges logarithmically with system size according to $|d\widetilde{GE}/d\beta|_{\beta=\beta_{m}}\approx \kappa \ln{N}$, where $\kappa\approx0.836$. Figure \ref{fig:3d} shows the convergence of $\beta_{m}$ to $\beta^*$. The y-interception $\beta_{m}(\infty)\approx 0.439$ is a reasonable approximation of analytical result $\beta^* \approx 0.441$. The inset in Fig. \ref{fig:3d} points out that the convergence to $\beta^*$ has a relation $|\beta^* - \beta_{m}|\sim N^{-\gamma}$ with the exponent $\gamma = 0.56$.

It is useful to sum up the results before closing this section. By exploiting the approach of quantum-classical mapping we were able to show that $GE$ and $\widetilde{GE}$ have classical correspondence in terms of expected total energy and energy-energy correlations of the Ising model. We find that both $GE$ and $\widetilde{GE}$ are monotonic decreasing functions of coupling with non-maximum value at the critical point. Indeed, they have large values in the topological phase where long-range entanglement exists.
However, $GE$ and $\widetilde{GE}$ are sensitive to the critical point in a sense that their first-order derivative diverges at $\beta^{*}$ in the thermodynamic limit (peak at $\beta_{m}$ for finite system) and hence they can be regarded as reasonable quantities to probe criticality. As a by-product, nonmaximality of $GE$ at the critical point in the absence of symmetry breaking adds to the evidence that a connection between symmetry breaking and $GE$ may exist \cite{effect,rad18,montakhab}.
As a next step, one might ask if there is another quantity which peaks at the critical point of TQPT? The following section aims to answer this question.
\section{conditional global entanglement}\label{sec3}
\begin{figure}[t]
\centering
\includegraphics[width=90mm,height=60mm,angle=0]{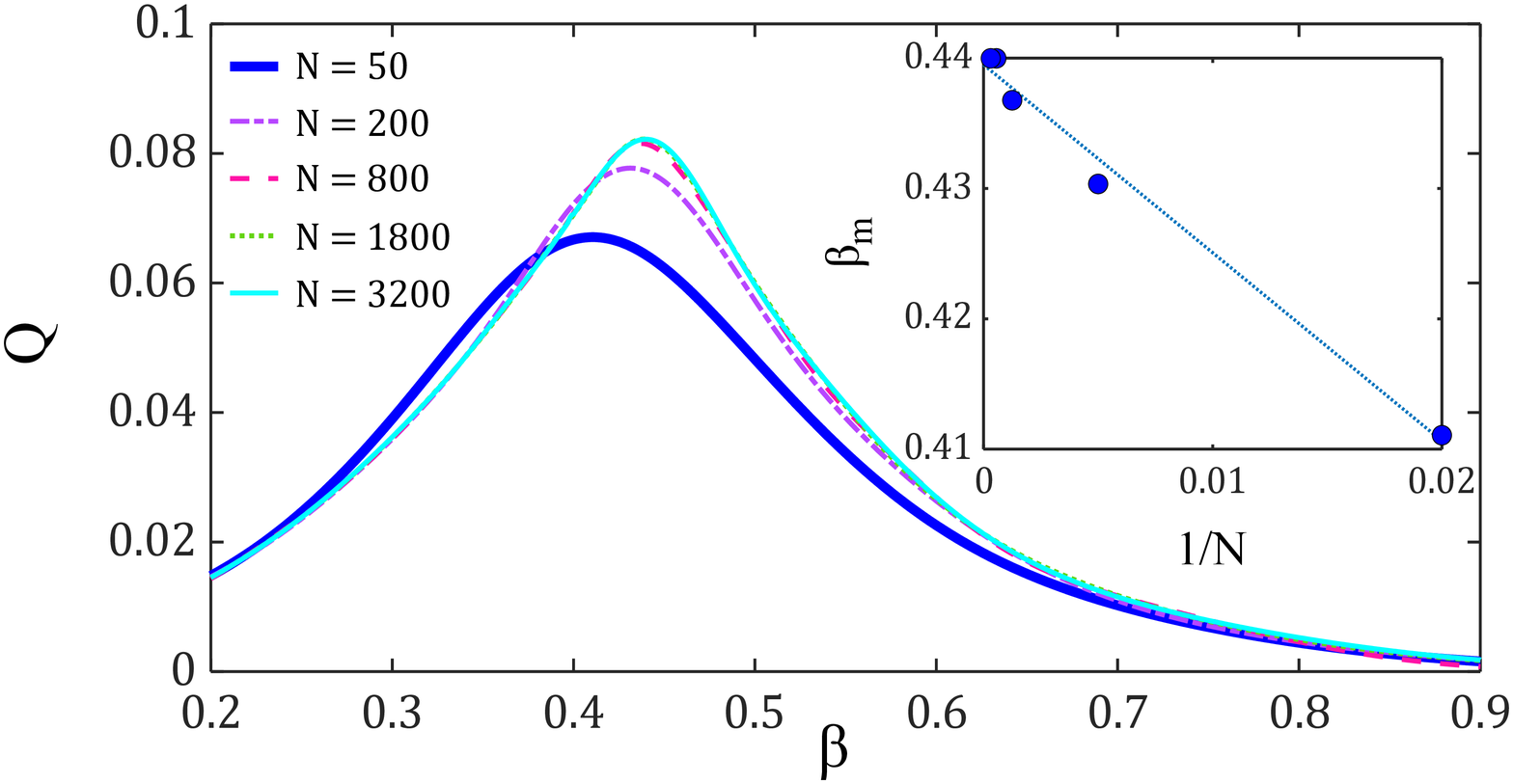}
\caption{(Color online) Conditional global entanglement $Q$ as a function of $\beta$ (coupling) for different system sizes, obtained by numerical simulations of the classical Ising model. The inset shows that $\beta_{m}(\infty)= 0.439$. The data for larger system sizes coincide at the present resolution.} \label{fig:4}
\end{figure}
As shown in the previous section, neither $GE$ nor $\widetilde{GE}$ peaks at the critical point but rather they are monotonic functions of $\beta$. We conjecture that the combination of $\widetilde{GE}-GE$ removes the effect of long-range entanglement in the topological phase. On the other hand, Since the combination of $\widetilde{GE}-GE$ is still a multipartite entanglement measure, we expect that it becomes maximum at the critical point.
However, $\widetilde{GE}-GE$ is a meaningful quantity which we explain below.

In quantum information theory, a useful quantity called quantum conditional entropy of a composite system with two components A and B is obtained through
\begin{equation}\label{C8}
S(A|B)\equiv S(A, B)-S(B),
\end{equation}
where $S(A, B)$ and $S(B)$ are the von Neumann entropy of the whole system and subsystem $B$, respectively. On the other hand, as already mentioned, $GE$ and $\widetilde{GE}$ have been regarded as entropy functions or mean linear entropy of one-qubit and two-qubit reduced density matrices. Therefore, if we consider two qubits $a$ and $b$ as a composite system, $\widetilde{GE}$ plays the role of the average of $S(A,B)$ and $GE$ plays the role of the average of $S(B)$. In this way, the difference between $\widetilde{GE}$ and $GE$ denoted by $Q$ is a quantity similar to the conditional entropy that we call the conditional global entanglement:
\begin{equation}\label{C3}
\begin{aligned}
Q(\hat \rho)&=\widetilde{GE}(\hat \rho)-GE(\hat \rho)
\\
&=\frac{1}{3}\frac{\langle E\rangle ^{2}}{N^2}-\frac{2}{3N(N-1)}\sum_{(ij)}\langle E_{i}E_{j}\rangle^{2}.
\end{aligned}
\end{equation}

We first seek to investigate the behavior of $Q$ analytically. When $T\rightarrow \infty$, there is no correlation between spins in the Ising model, so $\langle E\rangle=0$ and $\langle E_{i}E_{j}\rangle=0$. As a result $Q=0$. While for $T=0$, $\langle E\rangle=-N$ and $\langle E_{i}E_{j}\rangle=1$, and consequently Q becomes zero as well. As shown by \cite{kad,Hecht} the energy-density-energy-density correlation function of links $i$ and $j$, with typical distance $r_{ij}$, near the critical point in the classical Ising model is
\begin{equation}\label{C4}
f_{EE}(\epsilon,r_{ij})=\langle E_{i}E_{j}\rangle-\langle E_{i}\rangle\langle E_{j}\rangle\approx\frac{e^{-2\epsilon r_{ij}}}{r^2_{ij}},
\end{equation}
where $\epsilon=(4/T)|T-T_{c}|/T_{c}$ is defined near the critical temperature $T_{c}$. Recall that $\langle E\rangle / N = \langle E_{i}\rangle$. We expect that the leading contribution to the summation in Eq. (\ref{C3}) arises from typical $r_{ij}$'s. Hence, by combining Eq. (\ref{C4}) and Eq. (\ref{C3}), we can write $Q$ as
\begin{equation}\label{C5}
\begin{aligned}
Q&\approx\frac{1}{3}(\frac{\langle E\rangle^{2}}{N^2}-\frac{\langle E\rangle^{4})}{N^4})\\
&-\frac{2}{3N(N-1)}\sum_{(ij )}(\frac{e^{-4r_{i,j}\epsilon}}{r^4_{ij}}+2\frac{e^{-2r_{ij}\epsilon}}{r_{ij}^2}\frac{\langle E\rangle^{2}}{N^2}).
\end{aligned}
\end{equation}
Despite the first two terms of Eq. (\ref{C5}), which are constant, leading terms in the summation decrease with the increase of $r_{ij}$, such that in case $r_{ij}\rightarrow\infty$, they become zero. Near the critical point and for typical $r_{ij}$, these terms vanish exponentially, while at the critical point they decrease as power law but they are still relatively small in comparison with constant terms. Hence we estimate $Q$ in the thermodynamic limit as
\begin{equation}\label{C6}
Q\approx\frac{1}{3}(\frac{\langle E\rangle^{2}}{N^2}-\frac{\langle E\rangle^{4}}{N^4}).
\end{equation}
By differentiating with respect to $\langle E\rangle$ we obtain
\begin{equation}\label{C7}
\frac{\partial Q}{\partial\langle E\rangle}=0\longrightarrow(\frac{\langle E \rangle}{N})\approx\frac{1}{\sqrt{2}}.
\end{equation}
\\
On the other hand, as shown by \cite{coy}, the average total energy per link for the 2D square Ising model at critical point and in the thermodynamic limit is $1/\sqrt{2}$, which is in correspondence with the value obtained by Eq. (\ref{C7}). This result shows that $Q$ becomes maximum at criticality with the value  $Q(T_{c})=Q_{max}\approx0.083$.

We then turn to numerical calculations in order to support our
analytical argument. We have shown our results for different
system sizes up to $N=3200$ in Fig. \ref{fig:4}. Q
reaches its maximum value, $\approx0.082$, near criticality, which is in
line with analytical approximation. The inset shows the convergence of $\beta_m$ to $\beta^*$, where $\beta_{m}(\infty)\approx {0.439}$.
\\
\section*{Concluding remarks}
Quantum topological phases have attracted much attention over the
past two decades partly due to their potential for memory resource
in quantum computing. The associated phase transition to such
phases are also quite interesting because they do not possess the
usual symmetry-breaking mechanism associated with the local order
parameter in standard critical phenomena. However, topological
phase transitions are associated with long-range correlations
which underlie scaling properties of critical phenomena.
Multipartite entanglement has been used by various authors to
study and characterize many features of quantum phase transitions
in the past.  Such studies are partly motivated by the picture
that, in the critical state, long-range correlations should lead
to a highly entangled state. We have therefore used
well-known multipartite measures of entanglement ($GE$ and $\widetilde{GE}$) in
order to characterize TQPT in a perturbed Kitaev Toric code model. We have
shown that the topological phase is a highly entangled state with
a sharp change (maximum first-order derivative) at the critical point.
The scaling properties show a logarithmic divergence of
entanglement susceptibility.  This is similar to finite-size
studies of well-known models which exhibit symmetry-breaking phase
transitions.  However, in the present model, due to the exact
mapping to the classical counterpart [see Eq. (\ref{C1})], $GE$ and $\widetilde{GE}$ do
not exhibit maximum value at the phase-transition point in the
thermodynamic limit. We have finally proposed a measure based on
quantum conditional entropy which is able to show maximum value at
the critical point.  Whether (or not) such a measure is able to
exhibit maximum value at the critical point in other models is an
interesting avenue for future research. In particular, we notice that the ground state of Toric code is a loop fluctuating state and the nonlinear perturbation plays the role of  a string tension. Therefore, we expect that our conceptual arguments can be extended to other string-net condensed states \cite{string-net} in the presence of the string tension even if other forms of perturbation are used in the role of the string tension. Further investigations are needed in order to establish whether or not our expectation is met with supporting results.
\section*{Acknowledgement}
We would like to thank A. Ramezanpour and A. Bayat for fruitful discussions.

\end{document}